\documentclass[twocolumn,prl,amsfonts,showpacs,epsfig,floats,superscriptaddress,floatfix]{revtex4}
\usepackage{graphicx}
\usepackage{dcolumn}
\usepackage{bm}
\usepackage{longtable}
\usepackage{amsmath}

\begin{document}
\bibliographystyle{apsrev}


\title{Effects of QCD phase transition on gravitational radiation \\
from two-dimensional collapse and bounce of massive stars}%

\author{Nobutoshi Yasutake}
\email{yasutake@gemini.rc.kyushu-u.ac.jp}
\affiliation{Department of Physics, Kyushu University, Fukuoka 810-8560, Japan}
\author{Kei Kotake}
\affiliation{Division of Theoretical Astronomy, National Astronomical Observatory of Japan, 2-21-1 Osawa, Mitaka, Tokyo 181-8588, Japan}%
\author{Masa-aki Hashimoto}
\affiliation{Department of Physics, Kyushu University, Fukuoka 810-8560, Japan}
\author{Shoichi Yamada}
\affiliation{Science and Engineering, Waseda University, 3-4-1 Okubo, Shinjuku, Tokyo 169-8555, Japan}
\affiliation{Advanced Research Institute for Science and Engineering, Waseda University, 3-4-1 Okubo, Shinjuku,Tokyo, 169-8555, Japan}%

\date{\today}

\begin{abstract}
We perform two-dimensional, magnetohydrodynamical core-collapse simulations of 
massive stars accompanying the QCD phase transition. 
 We study how the phase transition affects the gravitational waveforms 
 near the epoch of core-bounce. 
As for initial models, we 
change the strength of rotation and magnetic fields. 
Particularly, the degree of differential rotation in the iron core (Fe-core) is changed arametrically. As for 
the microphysics, we adopt a phenomenological equation of state above 
the nuclear density, including two parameters to change the 
hardness before the transition. We assume the first order 
phase transition, where the conversion of bulk nuclear matter to a chirally 
symmetric quark-gluon phase is described by the MIT bag model. 
Based on these computations, 
we find that the phase transition can make the maximum amplitudes larger up to $\sim$ 10 percents than the ones without the phase 
transition.
On the other hand, the maximum amplitudes become smaller up to $\sim$ 10 percents owing to the phase transition, when
the degree of the differential rotation becomes larger.
 We find that even extremely strong magnetic fields $\sim 
10^{17}$ G in the protoneutron star do not affect these results.
\end{abstract}

\pacs{Valid PACS appear here}
\maketitle

\section{\label{sec:level1}Introduction}
It was presented that quark matter may exist in the universe~\cite{witten}.
The existence is now supposed to be inside the core of neutron stars or 
inferred to be bare quark stars, where several observational signals have been 
suggested ~\cite{drake,mira,li99sax,Xu05}. 
Much attention has also been paid to explore the relevant astrophysical 
phenomena in such sites ~\cite{bombaci04,drago,grigorian,page,yasutake}.
Related to the newer observation of PSR J0751+1807, skyrmion stars 
have been proposed to explain the large mass ($\sim2.1~M_{\odot}$) 
of a compact object~\cite{jai06}.

It has been presented that the quark matter might appear during supernova 
explosions~\cite{gentile,takahara}
 or the transition of a neutron star to a quark star~\cite{berezhiani}.
Current supernova studies demonstrate that the stellar collapse of stars below  
$\sim 25M_{\odot}$ in the main sequence stage leads to the formation of neutron stars, while in the case of 
more massive stars, to the formation of the black hole \cite{heger03}. 
In the latter case, quark matters might appear because the central density 
could exceed the density of the QCD phase transition.
It is worth mentioning here that supernova models with the 
phase transition are important because
 the vast release of the gravitational energy at the transition could be 
responsible for some classes of long-duration gamma-ray bursts 
\cite{yasutake}. In this connection, another energy source has been 
recently proposed as {\it quark nova} \cite{ouyed,ouyed05}. 

 It should be noted that the uncertainty of the equation of state (EOS) is always a big problem in 
the research of core-collapse supernovae with the phase transition. 
We might study such microphysical phenomena in the core of the star directly 
from the detections of the gravitational waves.
Currently, the gravitational astronomy is now becoming a reality. In fact, the 
ground-based laser interferometers such as TAMA300 \cite{tama,tamanew} and 
the first LIGO \cite{firstligo,firstligonew} are beginning to 
take data at sensitivities where astrophysical
events are predicted. For the detectors including GEO600 and VIRGO, core-collapse supernovae
especially in our Galaxy, have been supposed to be the most plausible
sources of gravitational waves (see reviews, for example, \cite{new,kotake_rev}). 

So far, there has been extensive 
work devoted to studying gravitational radiation in the context of rotational 
\cite{monchmeyer,yamada1995,zweg,dimmel,fry,kotakegw,shibaseki,ott} and magnetorotational 
\cite{kotake,obergaulinger} core collapse supernovae.
On the other hand, there are a very few simulations concerning the effects 
of the phase transition on the gravitational signals.
 Lin and his collaborators recently presented 
a three-dimensional simulation with the use of the polytropic equation 
of state in baryonic phase and MIT bag model in quark phase. They investigated
how the delayed collapse of a rapidly rotating neutron star induced by the 
phase transition long after its formation, could produce the
 gravitational waveforms ~\cite{lin}. They demonstrated that the waveforms 
can be characterized by the damping timescale of the core and the induced core 
oscillation frequency. Furthermore, they pointed that the 
energy release in the form of the gravitational radiation owing to the 
transition could be less than $\sim 10 \%$ of the gravitational binding energy.  

In the present paper, we also pay attention to the effects of the
phase transition on the gravitational waveforms, not at the epoch 
considered above, but at the moment of core-bounce during the gravitational 
collapse of the massive stars. 
For this purpose, we perform the two-dimensional  magnetohydrodynamic (MHD)
 simulations of the supernova cores accompanying the phase transition. 
To treat the QCD phase transition, 
we employ a MIT bag model. We construct the initial models by 
changing the strength of rotation and magnetic fields, and the degree of 
differential rotation parametrically. We also vary the hardness of EOS in the 
baryonic phase in a parametric manner. In this paper, 
we investigate the relation between the phase transition and the change 
in the gravitational signals systematically.
 
This paper is organized as follows.
In sec.~II we outline the numerical methods, input physics, the initial models,  and the numerical methods for the gravitational waveforms.
In Sec.~III, we present numerical results.
Sec.~IV is devoted to the conclusion.
\section{Numerical method and initial models}
\subsection{\label{sec:level2}Numerical method}
The numerical method for MHD computations employed in this paper is 
based on the ZEUS-2D code \cite{stone}, (see \cite{kotake} for details 
 of the application to the core-collapse simulations).
In the following equations, geometric units are used, $G=c=\hbar=1$.
 The basic MHD equations are written as follows, 
\begin{eqnarray}
\cfrac{D\rho}{Dt}\hspace{-2.5mm}& &   + \rho \bm{\nabla} \cdot \bm{v}  = 0,
\label{continuity}
\\
\rho\cfrac{D\bm{v}}{Dt}  =
-\bm{\nabla} P\hspace{-2.5mm}& & -  \rho \bm{\nabla} \Phi_{\textmd{eff}} +\cfrac{1}{4\pi}(\bm{\nabla}\times\bm{B})\times\bm{B},
\label{EOM}
\\
\rho\cfrac{D}{Dt} \left( \cfrac{e}{\rho} \right) \hspace{-2.5mm}& &  = -P\bm{\nabla}\cdot\bm{v},
\label{induction}
\\
\cfrac{\partial \bm{B}}{\partial t} \hspace{-2.5mm}& & =  \bm{\nabla} \times (\bm{v} \times \bm{B}),
\\
\Delta \Phi \hspace{-2.5mm}& & = 4 \pi \rho.
\label{poisson}
\end{eqnarray}
Here, $\rho,P,\bm{v},e$, $\bm{B}$, $\Phi$, and ${D}/{Dt}$,
 are the density, pressure, velocity, internal energy density, magnetic field, gravitational potential,
 and the Lagrangian derivative, respectively. 
In addition to the previous version
 \cite{kotake}, we newly take into account 
the general relativistic correction to the Newtonian gravity 
because it affects the waveform of the gravitational wave 
at the moment of the phase transition and 
the subsequent contraction of the core.
The effective potential $\Phi_{\textmd{eff}}$ in Eq. (\ref{EOM}) includes the
 correction \cite{marek}, which is defined to be 
\begin{eqnarray*}
\Phi_{\textmd{eff}} &=& \Phi + \delta \Phi_{\textmd{TOV}},
\end{eqnarray*}
where
\begin{eqnarray*}
\delta \Phi_{\textmd{TOV}} &=& \Phi_{\textmd{TOV}} - \Phi_{\textmd{S}}.
\end{eqnarray*}
Here, $\Phi_{\textmd{TOV}}$ is the gravitational potential which is 
constructed using the 
Tolman-Oppenheimer-Volkoff equation. $\Phi_{\textmd{S}}$ is the spherical 
Newtonian gravitational potential.
Albeit simple, this method has been demonstrated to reproduce well 
the results of the general relativistic calculations ~\cite{marek}.

\subsection{\label{sec:level2}Equation of State}

To describe the QCD phase transition in the supernova cores, we follow the
method adopted in Ref.~\cite{gentile}.
The transition to a hadronic phase is modeled by adding the MIT 
bag constant $B$ to the energy density. 
This phenomenological term lowers the pressure of the quark phase, where a transition to a confined phase occurs.

The EOS in the quark phase is written as :
\begin{eqnarray}
\rho \hspace{-0.5mm} & & \hspace{-3mm}= \cfrac{9}{4} \left( \cfrac{3\pi^2}{N_f} \right)^{1/3} \left[ 1+\cfrac{2\alpha_s}{3\pi}\right]n_b^{4/3}+B,
\label{ro}
\\
P \hspace{-0.5mm} & & \hspace{-3mm} = \cfrac{3}{4} \left( \cfrac{3\pi^2}{N_f} \right)^{1/3} \left[ 1+\cfrac{2\alpha_s}{3\pi}\right]n_b^{4/3}-B,
\label{P}
\\
\mu_b \hspace{-0.5mm} & & \hspace{-3mm} = 3\left( \cfrac{3\pi^2}{N_f} \right)^{1/3} \left[ 1+\cfrac{2\alpha_s}{3\pi}\right]n_b^{1/3}.
\label{mu}
\end{eqnarray}
Here, $\rho$ and $\mu_b$ are the energy density and the mean baryon chemical potential, respectively. We assume that $\mu_b$ is three times that of the quark chemical potential.
The number of quark flavors $N_f=3$, and $n_b$ is the mean baryon number density which is assumed  to be the one third of the quark number density. 
Using (\ref{ro}) and (\ref{P}), we get :
\begin{eqnarray*}
\rho  = 3P+4B.
\end{eqnarray*}

We assume that the first order phase transition occurs during the collapse beyond
some critical density, which has been supported by a recent study about the quark-gluon plasma~\cite{glen92}.
We set the QCD coupling constant $\alpha_s=0$ in most models, which indicates that the coexistence phase between the
baryon and quark is the widest in the density region.
This choice helps us evaluate the effects of the phase transition maximally~\cite{gentile}.
In the baryon phase, we adopt the phenomenological EOS of BCK~\cite{bck}
 that includes two parameters of the incompressibility $K$ and
 the adiabatic index $\Gamma$ to express the degree of the hardness of matter above $\rho_0$.
The EOS of BCK is parametrized as follows,
\begin{eqnarray}
P = \cfrac{Kn_0}{9\Gamma} \left[ \left( \cfrac{n_b}{n_0} \right)^\Gamma - 1 \right].
\label{BP}
\end{eqnarray}
Here the chemical potential in the baryon phase is given as, 
\begin{eqnarray}
\mu = \cfrac{\rho+P}{n_b},
\label{Bmu}
\end{eqnarray}
here $n_0$ is the saturation number density ($n_0=0.17$~fm$^{-3}$), and $\rho_0 =m_u n_0$ ($m_u$ is the atomic mass unit).
Two parameters of $K$ and $\Gamma$ are shown in Table \ref{tab:pt}.
We note that recent measurements give the constraints on the incompressibility : $ 210 \leq K\leq 270$ MeV \cite{uchida,piekarewicz,vretenar}.
Therefore the adopted value of $K=220$ MeV is consistent with this measurement.
We furthermore take into account the parameter of $K=375$ MeV, which would correspond to an extreme value.
We add the pressures of electrons and photons to the EOS of BCK~\cite{gentile}.
Since we focus on the behaviour of the EOS above $n_0$, we use
the phenomenological EOS of Ref.~\cite{yamada} below $n_0$ in all models.

As for the criterion of the phase transition, we impose the Gibbs condition with respect to the pressure and chemical
potential to bridge the baryon and quark phase.
We assume that the transition starts at $\rho_1 = 5\times 10^{14}$~g~cm$^{-3}$ for the baryon phase in most models,
whose value lies between those adopted by Gentile et al~\cite{gentile}.
Then, the end point of the transition ($\rho_2$) and $B$ are determined analytically from the condition
of $P=P_1$, $\mu=\mu_1$, and $\alpha_s=0$ or $0.25$ for the quark phase : $\rho_2$ and/or $B(\rho_1,P_1,\mu_1,\alpha_s)$.
Thus, we obtain $\rho_2$, $B$,
 and $\mu_b$ for two parameters of $K$ and $\Gamma$ as shown in Table \ref{tab:pt}.
The bag constants used in this paper are consistent with the values used 
to investigate the structure of hybrid and quark stars \cite{heiselberg,alford}. 
 
\begin{table*}
\caption{\label{tab:pt} Physical quantities for EOS in hadron/quark phase equilibrium.
The values, $\alpha_s$, $\rho_1$, $\rho_2$, $B^{1/4}$, and $\mu_b$ are the QCD coupling constant, the critical density of the baryon phase, the density of the corresponding quark phase, the bag constant, and the mean baryon chemical potential,  respectively.
$K$ and $\Gamma$ are the incompressibility and the adiabatic index, which change the hardness of the baryon phase.}
\begin{ruledtabular}
\begin{tabular}{lccccccc}
Model & $\alpha_s$ & $\rho_1$ ($10^{14}$g~cm$^{-3}$) & $\rho_2$ ($10^{14}$g~cm$^{-3}$) & $B^{1/4}$(MeV) & $\mu_b$(MeV) & $K$(MeV) & $\Gamma$ \\
\hline
-Mh & 0 & 5 & 7.34 & 164.8 & 967 & 375 & 3 \\
-Mm & 0 & 5 & 7.01 & 163.8 & 949 & 220 & 3 \\
-Ms & 0 & 5 & 6.93 & 163.8 & 949 & 220 & 2.5 \\
-Mma & 0.25 & 5 & 6.00 & 157.3 & 953 & 220 & 3 \\
-Mm6 & 0 & 6 & 7.28 & 163.8 & 964 & 220 & 3 \\
\end{tabular}
\end{ruledtabular}
\end{table*}

While we get a reasonable value of $\mu_b\sim950$~MeV,
$\rho_1$ and $\rho_2$ are rather low compared to the suggestion from the QCD theory
 based on the lattice QCD calculations. It is predicted that
 the transition occurs at $\mu_b \geq 1000$~MeV and $\rho_2 \geq 10^{15}~\rm g~cm^{-3}$ corresponding to the density of a neutron star~\cite{ivanov}.
We note that the pressure is not constant during the phase transition, since
 the electron and the photon pressures are included.

\subsection{\label{sec:level2}Initial models}

\begin{figure}
\includegraphics[width=77mm]{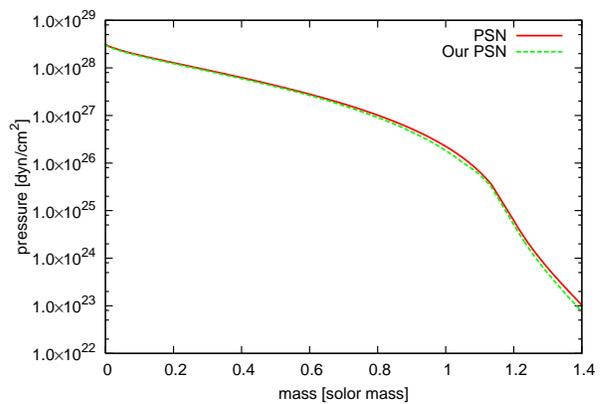}
\caption{\label{fig:compare} The difference of pressure between the original presupernova model~(PSN) and ours
(Our PSN), where the region of 
1.4 $M_\odot$ from the center is shown.}
\end{figure}

We adopt the presupernova model of 13~$M_{\odot}$ that has the iron core 
(Fe-core) of 1.2~$M_{\odot}$, in most models. 
The calculational area extends to 4000~km from the center with the mass of 1.4~$M_{\odot}$ included.
FIG.~\ref{fig:compare} shows the difference of the pressure between the original presupernova model and ours
whose EOS is described in Sec. II B.
We can see that our EOS is phenomenological but consistent with original one.
To examine the mass dependence of presupernova models,
we adopt an another presupernova model whose mass 
is 40~$M_{\odot}$ with the Fe-core of 1.9~$M_{\odot}$.
In this model, the calculational area of 4000~km corresponds to 2.4~$M_{\odot}$.

 Since the effects of the magnetic field and the
angular momentum distribution on the presupernova star are
 uncertain \cite{heger}, we make precollapse models 
from the nonrotating progenitor model by adding the angular
momentum and the magnetic field according to the following prescription.
For the initial angular velocity distribution, we adopt the $shell-type$ rotational law~\cite{heger}:
\begin{eqnarray*}
\Omega (r) &=& \Omega_0 \times \frac{R_0^2}{r^2+R_0^2},
\end{eqnarray*}
where $\Omega(r)$ is the angular velocity and $r$ is the radius. 
Both $\Omega_0$ and $R_0$ are model constants that prescribe the rotational law. 
We regard the model with $R_0=10^3$~km as uniformly rotating, since the radius of Fe core is $\sim 10^3$ km.
In most initial models, the initial magnetic field is assumed to be constant, $B_0$, which is poloidal and parallel to  
 the rotational axis in the computational domain.
For the only two models in Table \ref{tab:table01}, 
we assume that the initial magnetic field is purely toroidal:
\begin{eqnarray*}
B_\phi(r) &=& B_0 \times \frac{R_0^2}{r^2+R_0^2},
\end{eqnarray*}
where $B_{\phi}(r)$ is toroidal component of the magnetic fields 
and $B_0$ is a constant. The dominance of the toroidal fields to the 
poloidal ones is predicted by the recent stellar evolution calculation 
\cite{heger03,spruit}.
As is discussed in Sec. III, effects of the magnetic field distribution 
on the transition are found to be small.

We perform core-collapse simulations of 27 models which 
are given in Table \ref{tab:table01}.
The characters in the left hand side for each column indicate the 
initial condition concerning the
magnetic fields and rotations:
S~(spherical),
 u~(weak uniform rotation and weak magnetic field),
 U~(strong uniform rotation and strong magnetic field),
 R~(strong differential rotation),
 d~(strong differential rotation and strong magnetic filed),
 B~(strong differential rotation and strong magnetic field of the $shell-type$ law),
 D~(very strong differential rotation and strong magnetic field). 
If there is "40" in the model name, it means that the mass of the presupernava model is 40~$M_{\odot}$. 
The characters after the hyphen indicate the adopted EOSs.
The inclusion of the QCD phase transition is indicated by 'M': 
B~(BCK without the phase transition), and M~(MIT bag model with the phase transition).
The right hand side characters mean the hardness for the EOSs in the baryon phase or another choice in the transition parameters.
This hardness is expressed by the adiabatic index $\Gamma$ and the incompressibility $K$:
h~(hard ; $\Gamma=3,K=375$~MeV), m~(medium ; $\Gamma=3,K=220$~MeV),
and s~(soft ; $\Gamma=2.5,K=220$~MeV).
Two models u-Mma and d-Mma indicate that $\alpha=0.25$. The model u-Mm6 means that the critical density in baryon phase is $\rho_1 = 6 \times 10^{14}$~g~cm$^{-3}$. We do not calculate the case which corresponds to the model d-Mm, because the rotation of the core is so fast that the maximum density does not reach the critical density. 
\begin{table*}
\caption{\label{tab:table01}Model parameters. This table is divided into three groups : spherical (S- models), uniform rotation (u, 40u, U- models) and differential rotation  groups (d,  R, 40d, B, D- models). 
Note that the models with second capitals of ``B'' or ``M''  correspond to 
those without or with phase transition, respectively.} 
\begin{ruledtabular}
\begin{tabular}{lccccccccc}
Model & $\alpha_s$ & $\rho_1$ & $K$ & $\Gamma$ & $R_0$ & $T/|W|_{\textmd{ini}}$ & $E_{\textmd{m}}/|W|_{\textmd{ini}}$ & $\Omega_0$ & $B_0$ \\
& & ($10^{14}$ g cm$^{-3}$) & (MeV) &  & (km) & (\%) & (\%) & ($s^{-1}$) & (G) \\ \hline \hline
S-Bm& - & 5 & 220 & 3 & - & - & - & - &- \\
S-Mm & 0 & 5 & 220 & 3 & - & - & - & - & - \\ \hline \hline
u-Bh & - & 5 & 375 & 3 & $10^3$ & 0.1 & $10^{-3}$ & 2.3 & $6.7 \times 10^{10}$  \\
u-Mh & 0 & 5 & 375 & 3 & $10^3$ & 0.1 & $10^{-3}$ & 2.3 & $6.7 \times 10^{10}$ \\ \cline{2-10}
u-Bm & - & 5 & 220 & 3 & $10^3$ & 0.1 & $10^{-3}$ & 2.3 & $6.7 \times 10^{10}$ \\
u-Mma & 0.25 & 5 & 220 & 3 & $10^3$ & 0.1 & $10^{-3}$ & 2.3 & $6.7 \times 10^{10}$ \\
u-Mm6 & 0 & 6 & 220 & 3 & $10^3$ & 0.1 & $10^{-3}$ & 2.3 & $6.7 \times 10^{10}$ \\
u-Mm & 0 & 5 & 220 & 3 & $10^3$ & 0.1 & $10^{-3}$ & 2.3 & $6.7 \times 10^{10}$ \\  \cline{2-10}
u-Bs & - & 5 & 220 & 2.5 & $10^3$ & 0.1 & $10^{-3}$ & 2.3 & $6.7 \times 10^{10}$ \\ 
u-Ms & 0 & 5 & 220 & 2.5 & $10^3$ & 0.1 & $10^{-3}$ & 2.3 & $6.7 \times 10^{10}$ \\ \hline
40u-Bm & - & 5 & 220 & 3 & $10^3$ & 0.1 & $10^{-3}$ & 2.1 & $9.2 \times 10^{10}$ \\
40u-Mm & 0 & 5 & 220 & 3 & $10^3$ & 0.1 & $10^{-3}$ & 2.1 & $9.2 \times 10^{10}$ \\ \hline
U-Bm & - & 5 & 220 & 3 & $10^3$ & 0.5 & $10^{-1}$ & 5.3 & $ 6.7 \times 10^{11}$ \\ 
U-Mm & 0 & 5 & 220 & 3 & $10^3$ & 0.5 & $10^{-1}$ & 5.3 & $ 6.7 \times 10^{11}$ \\ \hline \hline
d-Bm & - & 5 & 220 & 3 & $10^2$ & 0.5 & $10^{-1}$ & 58.8 & $6.7 \times 10^{11}$ \\ 
d-Mma & 0.25 & 5 & 220 & 3 & $10^2$ & 0.5 & $10^{-1}$ & 58.8 & $6.7 \times 10^{11}$ \\
d-Mm & 0 & 5 & 220 & 3 & $10^2$ & 0.5 & $10^{-1}$ & 58.8 & $6.7 \times 10^{11}$ \\ \hline
R-Bm & - & 5 & 220 & 3 & $10^2$ & 0.5 & - & 58.8 & - \\
R-Mm & 0 & 5 & 220 & 3 & $10^2$ & 0.5 & - & 58.8 & - \\ \hline
40d-Bm & - & 5 & 220 & 3 & $10^2$ & 0.5 & $10^{-1}$ & 79.4 & $9.2 \times 10^{11}$ \\
40d-Mm & 0 & 5 & 220 & 3 & $10^2$ & 0.5 & $10^{-1}$ & 79.4 & $9.2 \times 10^{11}$ \\ \hline
B-Bm \footnote{For two models of B-Bm and B-Mm, the shell-type distribution of the magnetic field $B_{\phi}=B_{0}\times R_0^2/(r^2+R_0^2)$ is assumed.}
& - & 5 & 220 & 3 & $10^2$ & 0.5 & $10^{-1}$ & 58.8 & $1.3 \times 10^{14}$ \\
B-Mm & 0 & 5 & 220 & 3 & $10^2$ & 0.5 & $10^{-1}$ & 58.8 & $1.3 \times 10^{14}$ \\ \hline
D-Bm & - & 5 & 220 & 3 & 50 & 0.5 & $10^{-1}$ & 183.9 & $6.7 \times 10^{11}$ \\ 
D-Mm & 0 & 5 & 220 & 3 & 50 & 0.5 & $10^{-1}$ & 183.9 & $6.7 \times 10^{11}$ \\ 
\end{tabular}
\end{ruledtabular}
\end{table*}
\subsection{\label{sec:level2}Gravitational wave formulae from the rotating magnetized stellar cores}
To compute the gravitational waveforms from the rotating and magnetized stellar core, 
we follow the method of the quadrupole formula derived in \cite{kotake}. 
We describe the gravitational wave amplitude (GWA) $h_{ij}$ that is calculated by
\begin{eqnarray}
h^{\rm TT}_{ij}(R)=\cfrac{2}{R}\cfrac{d^2}{dt^2}I^{\rm TT}_{ij}(t-R),
\label{htt0}
\end{eqnarray}
where subscripts $i$ and $j$ take over $x$, $y$, and $z$. $t$ is the proper time and $R$ is the distance from the
source to an observer, respectively.
The superscript TT indicates the transverse traceless part of the metric. 
$I^{\rm TT}_{ij}$ is the reduced quadrupole moment, defined as
\begin{eqnarray*}
I^{\rm TT}_{ij}=\int \rho_*(x_ix_j-\cfrac{1}{3}x^2\delta_{ij} ) d^3x, 
\end{eqnarray*}
where $\rho_*$ is the total energy density including the 
contributions from the magnetic field,
\begin{eqnarray}
\rho_*=\rho+\cfrac{B^2}{8\pi},
\label{rho}
\end{eqnarray}
with $\rho$ being the matter density.
The amplitude (\ref{htt0}) is then transformed to the spherical coordinate as
\begin{eqnarray}
h^{\rm TT}=
h^{\rm TT}_{\theta \theta }=\cfrac{1}{8} \left( \cfrac{15}{\pi} \right)^{1/2}\sin^2{\alpha \cfrac{A^{E2}_{20}}{R}},
\label{httxx}
\end{eqnarray}
where $\alpha$ is the angle between the symmetry axis of the source and the direction to the observer and $A^{E2}_{20}$ is the second derivative of the radiative
quadrupole $M^{E2}_{20}$,
\begin{eqnarray}
A^{E2}_{20} = \frac{d^2}{dt^2}M^{E2}_{20}.
\label{ddmdtdt}
\end{eqnarray}
$A^{E2}_{20}$ consists of the following two terms:
\begin{eqnarray}
A^{E2}_{20} \equiv A^{E2}_{20,\textmd{quad}} + A^{E2}_{20,\textmd{Mag}}.
\label{A20}
\end{eqnarray}
Here, $A^{E2}_{20,\textmd{quad}}$ is the contribution from the matter.
The magnetic component in Eq.~(\ref{A20}) is decomposed into two terms:
\begin{eqnarray}
A^{E2}_{20,\textmd{Mag}} = A^{E2}_{20,\bm{j} \times \bm{B}}+ A^{E2}_{20,\rho_m},
\end{eqnarray}
where $A^{E2}_{20,\bm{j}\times \bm{B}}$ is
the contribution from the $\bm{j}\times\bm{B}$ part and $A^{E2}_{20,\rho_m}$  is that from the time derivatives of
the energy density of the electromagnetic field.
In consequence, Eq.~(\ref{httxx}) is composed of three terms :
\begin{eqnarray}
h^{\textmd{\rm TT}} = h^{\textmd{\rm TT}}_{\rm quad}+ h^{\textmd{\rm TT}}_{\bm{j}\times \bf{B}} + h^{\textmd{\rm TT}}_{\rho_m},
 \label{htt}
\end{eqnarray}
(see Ref. \cite{kotake} for details).
In the following, we assume that the distance from the observer ($R$) 
is 10 kpc in the direction of the equator ($\alpha=\pi/2$).

\section{Numerical results}
We summarize the physical quantities obtained from the numerical simulations 
in Table \ref{tab:result}. As a guide to see the effects of the phase 
transition on the maximum amplitudes, we prepare the following quantity, 
\begin{eqnarray}
\Delta |h^{\textmd{\rm TT}}|_{\textmd{max}}
 = \cfrac{|h^{\textmd{\rm TT}}_M|_{\textmd{max}}-|h^{\textmd{\rm TT}}_B|_{\textmd{max}}}{|h^{\textmd{\rm TT}}_B|_{\textmd{max}}},
\label{deltah}
\end{eqnarray}
where $|h^{\textmd{\rm TT}}_M|_{\textmd{max}}$ and $ |h^{\textmd{\rm TT}}_B|_{\textmd{max}}$ are
 the absolute value of gravitational wave amplitudes (GWAs) with and without the phase transition, respectively.

From the table,  we can see that the phase transition makes
the maximum GWAs larger by a few up to $\sim$ 10 percents for 
 the uniformly rotating models. 
On the other hand, the phase transition lowers the maximum GWAs
for the differentially rotating models.
If the rotational strength $T/|W|$ becomes very large, 
the effects of phase transition on gravitational waves become small. 
We explain these features in the following subsections, where
the changes in GWA from models to models are examined.
Furthermore, we perform the Fourier transformations for all models to show more comprehensive and systematic analysis of the gravitational waveforms.
\begin{table*}
\caption{\label{tab:result}Physical quantities for the models with and without the phase transition.
 $t_b ,\, t_{\textmd{fin}}$ and $\rho_{\textmd{max}}$
 are the bounce time, the final time of the calculation, and the maximum density. 
$T/|W|_{\textmd{fin}}$ and $E_{\textmd{m}}/|W|_{\textmd{fin}}$
 are the final values of $T/|W|$ and $E_{\textmd{m}}/|W|$, respectively. 
$|h^{\textmd{\rm TT}}|_{\textmd{max}}$ is the maximum GWA, and 
$\Delta |h^{\textmd{\rm TT}}|_{\textmd{max}}$ is defined by Eq.(\ref{deltah}).
Note that models with second capitals of ``B'', ``M''  corresponds to 
without and with phase transition, respectively.
$f_{1st}$ is the first peak frequency in the Fourier transformation of GWA.
}
\begin{ruledtabular}
\begin{tabular}{lcccccccc}
Model & $t_b$ & $t_{\textmd{fin}}$ & $\rho_{\textmd{max}}$ &
 $T/|W|_{\textmd{fin}}$  & $E_{\textmd{m}}/|W|_{\textmd{fin}}$ & $|h^{\textmd{\rm TT}}|_{\textmd{max}}$ & $\Delta |h^{\textmd{\rm TT}}|_{\textmd{max}}$ & $f_{1st.}$ \\
& (ms) & (ms) & (10$^{14}$g/cm$^3$) & (\%) & (\%) & ($10^{-20}$)& (\%) & (Hz) \\
\hline \hline
S-Bm & 75.5 & 126 & 6.95 & - & - & - & - & - \\
S-Mm & 75.5 & 146 & 8.84 & - & - & - & - & -\\
\hline \hline
u-Bh & 76.3 & 82.6 & 5.37 & 1.24 & 2.12 $\times 10^{-3}$ & 0.28 & - & 727 \\
u-Mh & 76.3 & 82.7 & 6.72 & 1.24 & 2.05 $\times 10^{-3}$ & 0.29 & +3.6 & 748\\
\cline{2-9}
u-Bm & 76.2 & 82.7 & 6.90 & 1.29 & 2.07 $\times 10^{-3}$ & 0.31 & - & 652 \\
u-Mma & 76.3 & 87.2 & 7.19 & 1.31 & 2.24 $\times 10^{-3}$ & 0.32 & +3.2 & 652\\
u-Mm6 & 76.2 & 83.4 & 8.05 & 1.27 & 2.24 $\times 10^{-3}$ & 0.33 & +6.4 & 838\\ 
u-Mm & 76.3 & 87.0 & 8.60 &1.33 & 4.27 $\times 10^{-3}$ & 0.34 & +9.6 & 875 \\
\cline{2-9}
u-Bs & 76.3 & 83.3 & 8.12 & 1.29 & 2.25 $\times 10^{-3}$ & 0.34 & - & 587 \\ 
u-Ms & 76.2 & 86.6 & 8.79 & 1.36 & 3.28 $\times 10^{-3}$ & 0.38 &+11.8 & 931 \\
\hline 
40u-Bm & 70.6 & 76.1 & 7.02 & 1.29 & 4.15 $\times 10^{-2}$  & 0.26 & - & 635(889) \\
40u-Mm & 70.7 & 79.4 & 8.64 & 1.39 & 5.42 $\times 10^{-2}$ & 0.28 & +7.6 & 762 \\ 
\hline 
U-Bm & 79.3 & 86.0 & 6.35 & 5.05 & 5.32 $\times 10^{-2}$ & 1.25 & - & 859 \\ 
U-Mm & 79.4 & 93.2 & 8.03 & 4.97 & 8.25 $\times 10^{-2}$ & 1.27 & +1.6 & 872 \\ 
\hline  \hline
d-Bm & 83.8 & 90.9 & 5.64 & 9.41 & 1.16 $\times 10^{-1}$ & 3.03 & - & 792 \\ 
d-Mma & 83.8 & 91.0 & 6.27 & 9.52 & 1.15 $\times 10^{-1}$  & 2.79 & $-$7.9 & 743 \\
d-Mm & 83.8 & 91.0 & 7.31 & 9.40 & 1.14 $\times 10^{-1}$ &  2.72 &$-$10.2 & 659 \\ 
\hline
R-Bm & 83.8 & 90.9 & 5.67 & 9.53 & - & 3.03 & - & 791 \\
R-Mm & 83.8 & 91.0 & 7.40 & 9.51 & - & 2.71 & $-$10.6 & 658 \\
\hline
40d-Bm & 84.1 & 89.9 & 5.81 & 12.8 & 1.33 $\times 10^{-1}$ & 2.98 & - & 562 \\
40d-Mm & 84.3 & 90.1 & 7.50 & 12.3 & 1.40 $\times 10^{-1}$ & 2.72 & $-$8.7 & 494 \\
\hline
B-Bm & 84.1 & 91.0 & 5.57 & 9.51 & 1.61 $\times 10^{-1}$ & 2.96 &- & 668 \\
B-Mm & 84.2 & 91.2 & 7.29 & 9.49 & 1.62 $\times 10^{-1}$ & 2.69 &$-$9.1 & 668 \\ 
\hline
D-Bm & 83.9 & 90.9 & 5.21 & 8.33 & 1.59 $\times 10^{-1}$ & 3.02 &- & 814 \\ 
D-Mm & 83.9 & 91.1 & 7.32 & 8.36 & 1.47 $\times 10^{-1}$ & 2.74 &$-$9.3 & 780\\ 
\end{tabular}
\end{ruledtabular}
\end{table*}

\subsection{\label{sec:level2}Effects of the phase transition and rotation}
\begin{figure*}
\includegraphics[width=75mm]{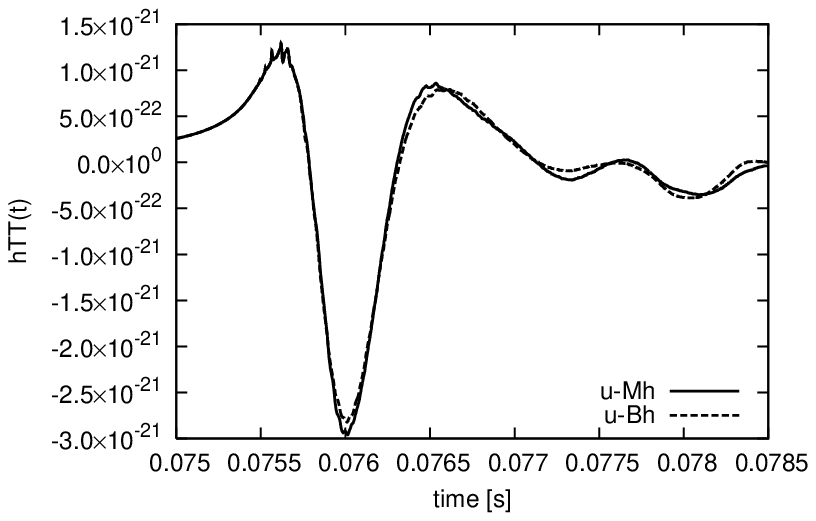}
\hspace*{5mm}
\includegraphics[width=75mm]{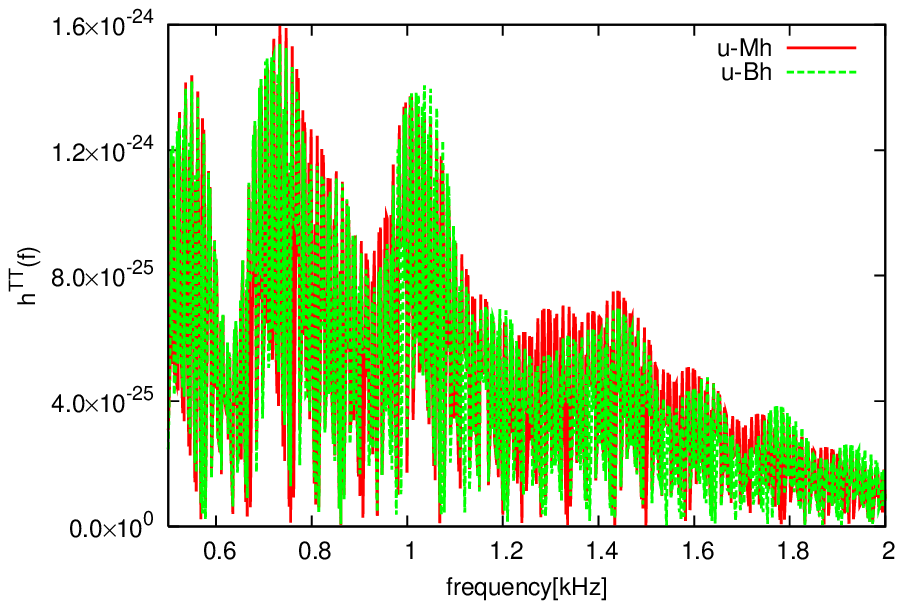}
\caption{\label{fig:GWh} GWAs as a function of time in u-Mh and u-Bh models (left)
and the corresponding Fourier transformations (right). }
\end{figure*}

\begin{figure*}
\includegraphics[width=75mm]{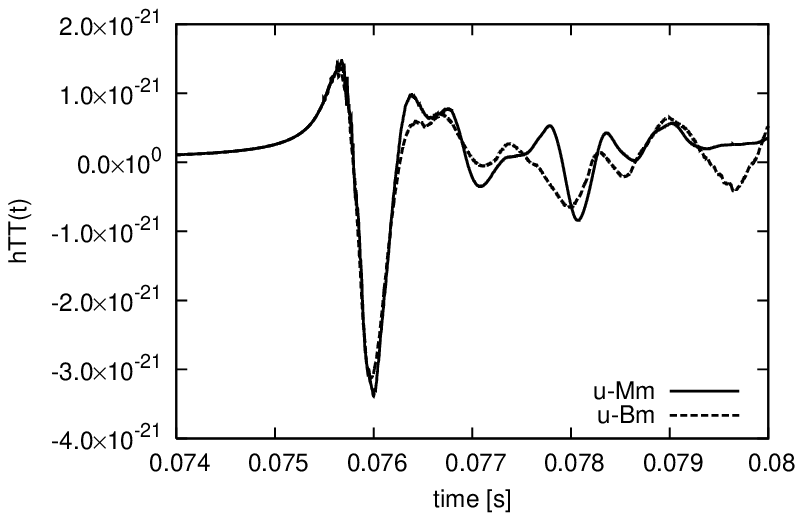}
\hspace*{5mm}
\includegraphics[width=75mm]{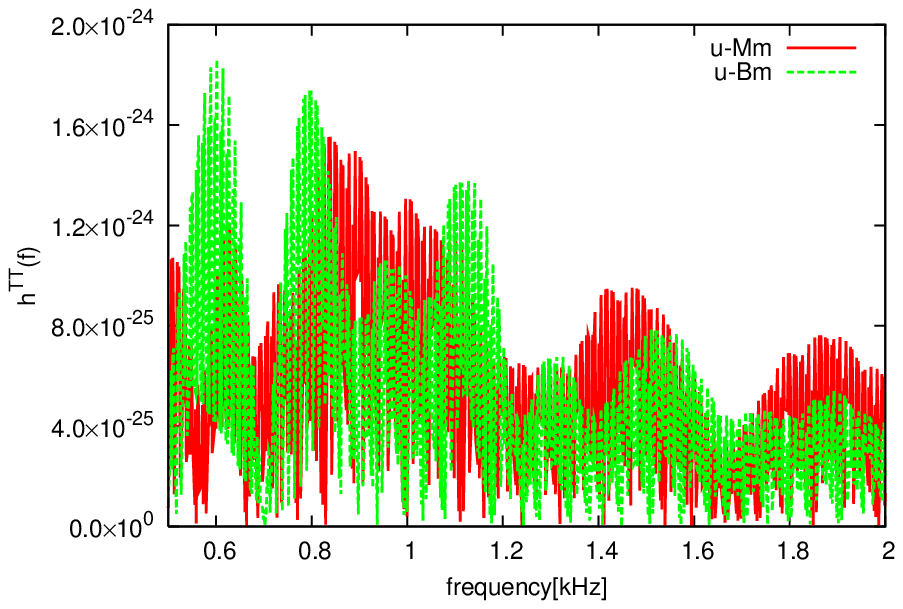}
\caption{\label{fig:GWm} Same as FIG.~\ref{fig:GWh} but for u-Mm and u-Bm.}
%
%
\includegraphics[width=75mm]{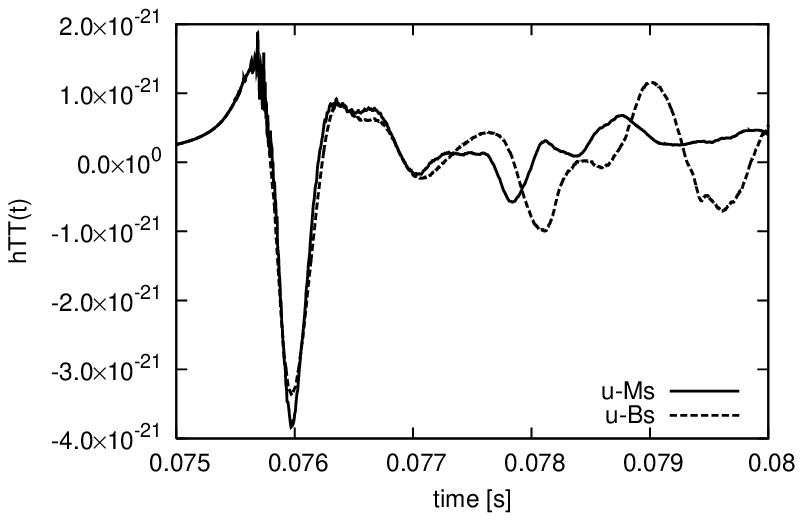}
\hspace*{5mm}
\includegraphics[width=75mm]{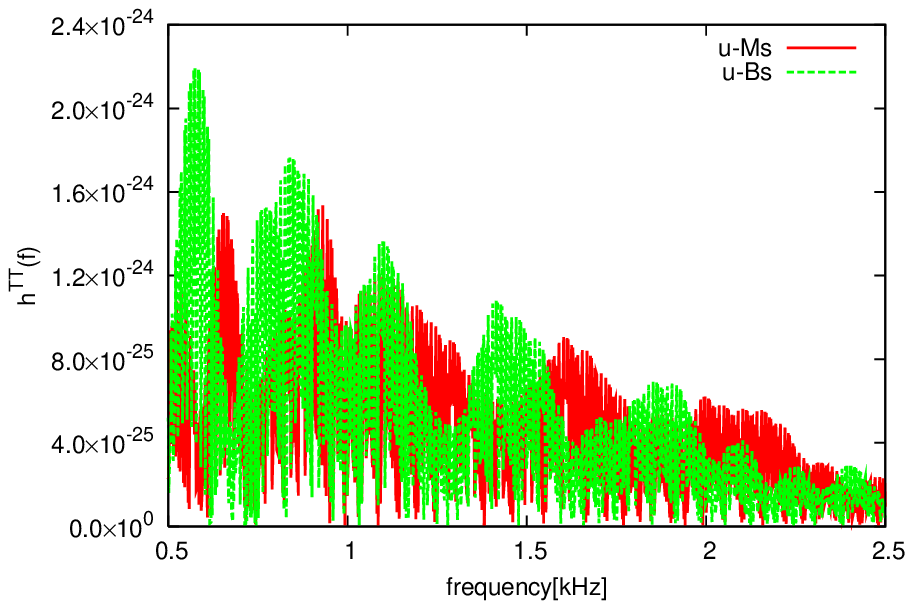}
\caption{\label{fig:GWs} Same as FIG.~\ref{fig:GWh} but for u-Ms and u-Bs. }
%
%
\includegraphics[width=70mm]{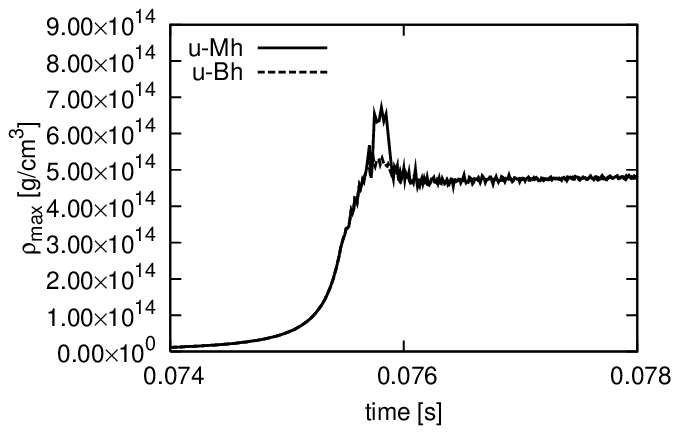}
\hspace*{5mm}
\includegraphics[width=70mm]{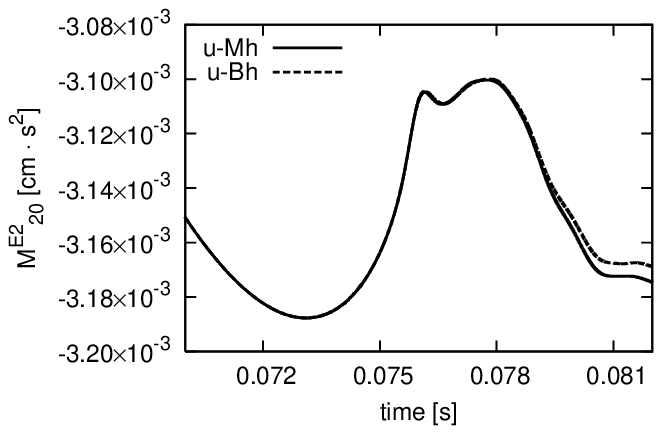}
\caption{\label{fig:GWQh} Maximum density and radiative quadrupole moment of the model u-Mh and u-Bh. } 
%
\includegraphics[width=70mm]{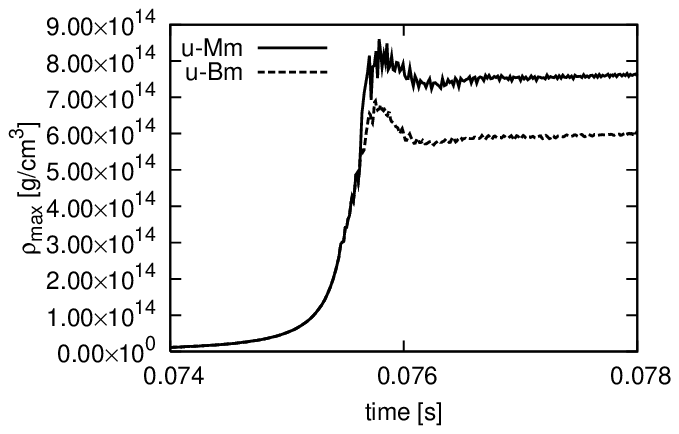}
\hspace*{5mm}
\includegraphics[width=70mm]{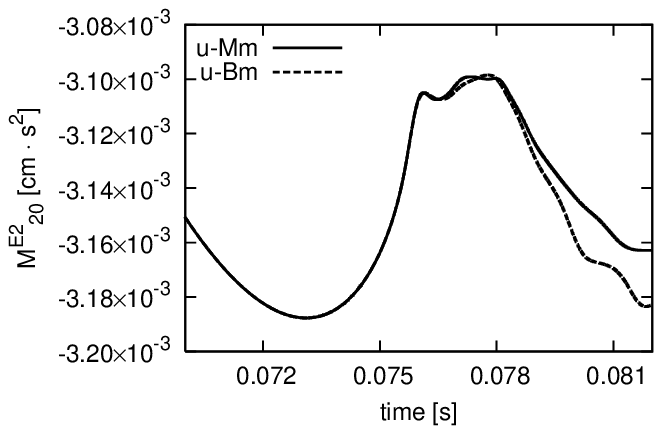}
\caption{\label{fig:GWQm} Same as FIG.~\ref{fig:GWQh} but for u-Mm and u-Bm. } 
\end{figure*}

We will first show the effects of the phase transition on GWA
 for the uniformly rotating models.
We find that GWAs are larger from a few percents to about ten percents
by considering the phase transition for all uniformly rotating models as seen
in Table~\ref{tab:result} and FIGs.~\ref{fig:GWh}-\ref{fig:GWs}.
\begin{figure*}
\includegraphics[width=70mm]{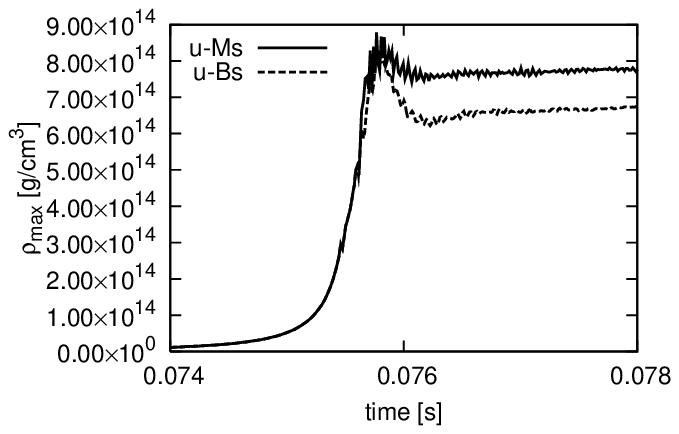}
\hspace*{5mm}
\includegraphics[width=70mm]{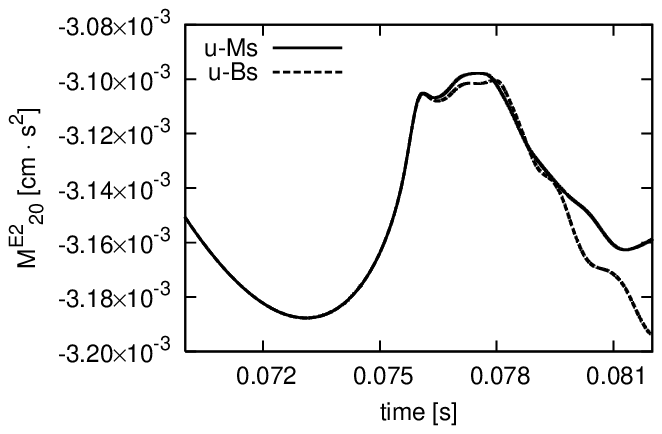}
\caption{\label{fig:GWQs} Same as FIG.~\ref{fig:GWQh} but for u-Ms and u-Bs. }
\end{figure*}
\begin{figure*}
\includegraphics[width=7.3cm]{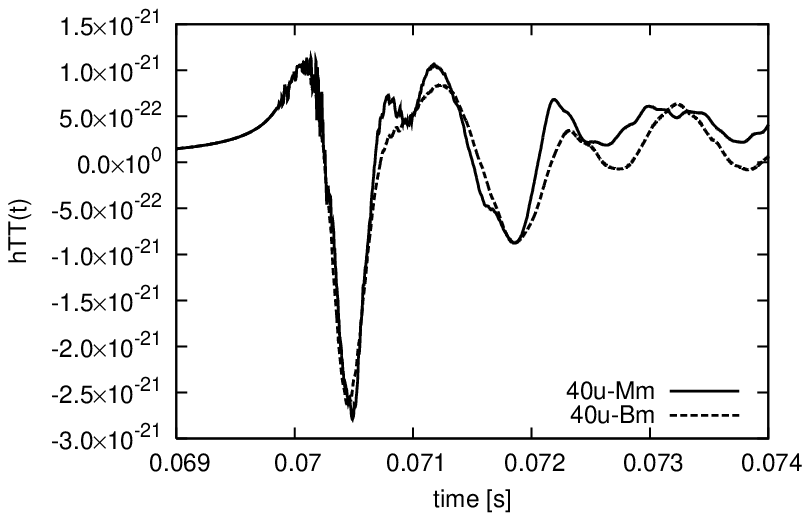}%
\hspace*{10mm}
\includegraphics[width=7.3cm]{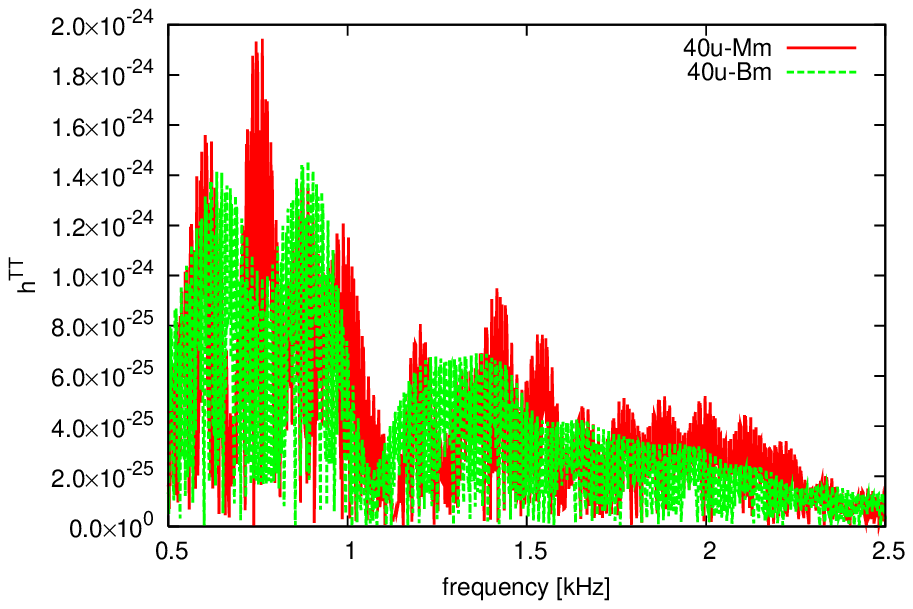}%
\caption{\label{fig:40u} Same as FIG.~\ref{fig:GWh} but for models of 40u-Mm and 40u-Bm.}
%
%
\includegraphics[width=70mm]{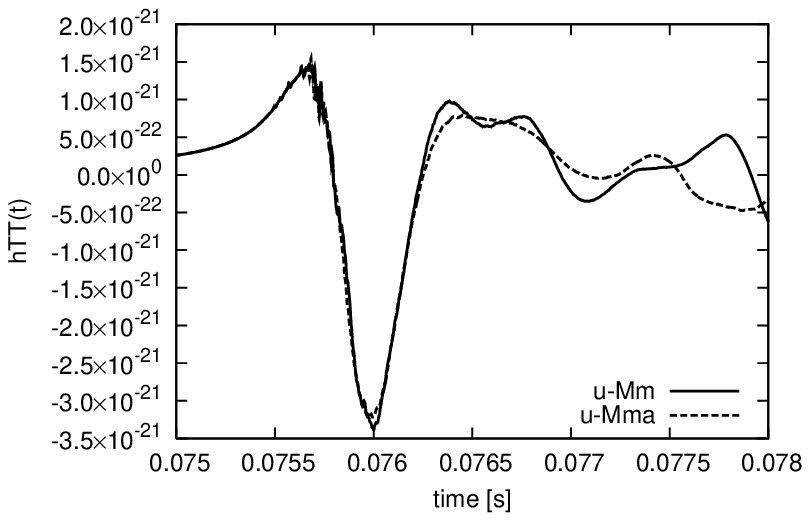}
\hspace*{10mm}
\includegraphics[width=70mm]{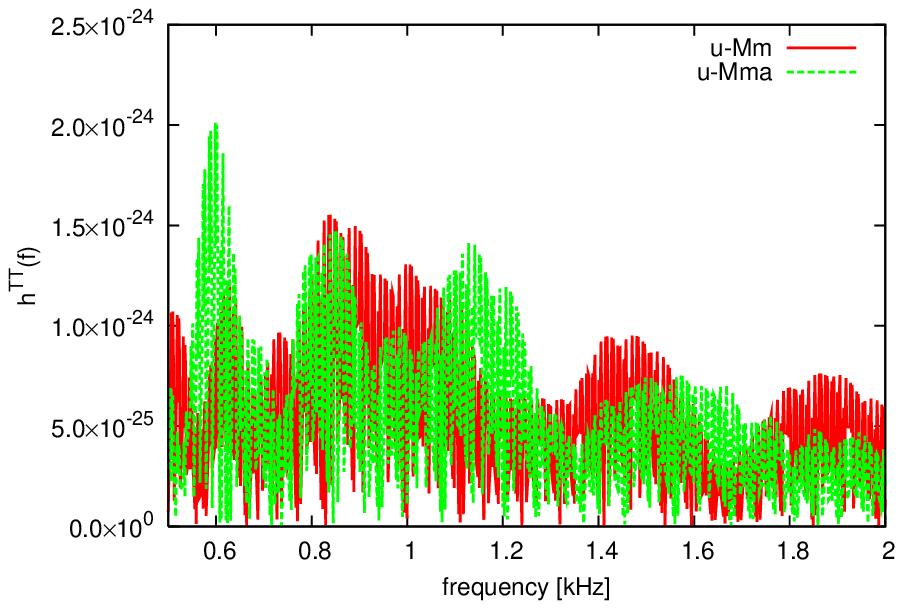}
\caption{\label{fig:GWua} Same as FIG.~\ref{fig:GWh} but for the models of u-Mm and u-Mma. }
%
%
\includegraphics[width=70mm]{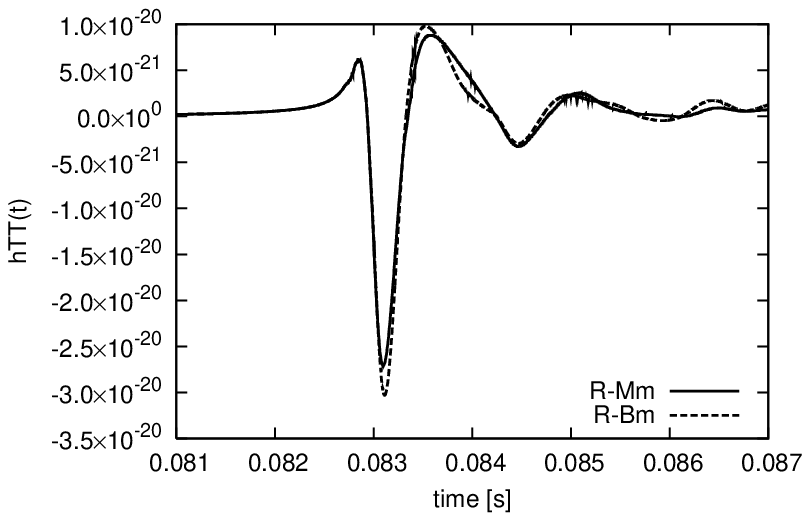}
\hspace*{10mm}
\includegraphics[width=70mm]{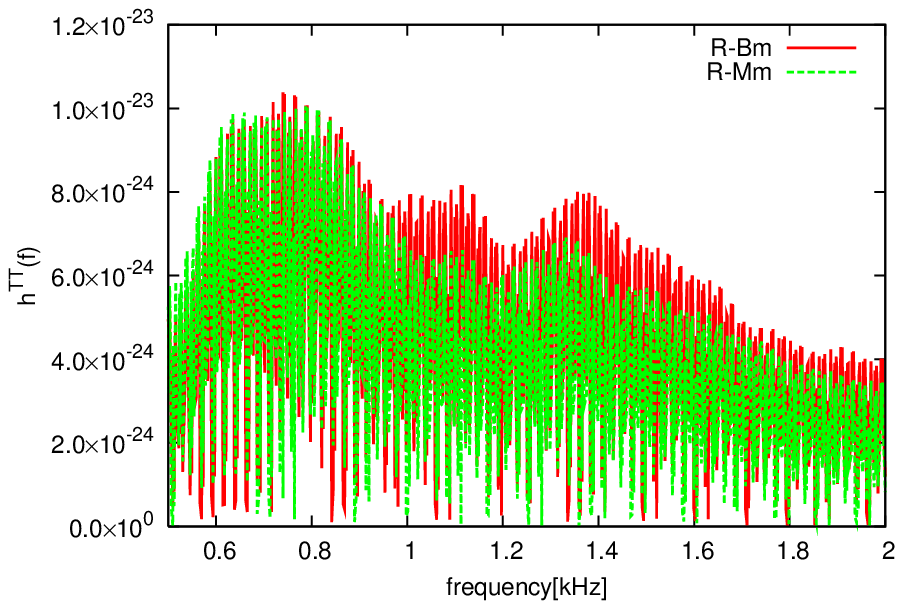}
\caption{\label{fig:GWR} Same as FIG.~\ref{fig:GWh} but for the models of R-Mm and R-Bm. }
\end{figure*}
 As the initial rotational strength $T/|W|_{\textmd{ini}}$ increases, the maximum GWA, $|h^{\rm TT}|_{\textmd{max}}$ becomes large.
 However the effect of the phase transition on the maximum GWA decreases as seen from $\Delta|h^{\textmd{TT}}|_{\textmd{max}}$.
 In the right panels of FIGs.~\ref{fig:GWh}-\ref{fig:GWs}, we can see that the frequency region with the phase transition shifts to the higher one as a whole, because the increased density during the collapse tends to the higher frequency ~\cite{lin}.
As for the effects of the equation of state, we find that the larger gravitational wave is radiated for the soft EOS, regardless of the phase transition.
Comparing the models of hard, medium and soft EOS in Table \ref{tab:result},
 this tendency is recognized to be independent on whether the softness originates from $\Gamma$ or $K$.

To understand these results, we give an order-of-magnitude estimation 
of the quadrupole formula \cite{yamada,kotake}.
The gravitational amplitude is proportional to the second time derivative of the radiative quadrupole
(see Eqs. (\ref{httxx}) and (\ref{ddmdtdt})).
The second time derivative in Eq.~(\ref{ddmdtdt}) could be replaced by the reciprocal of the square of the time,
 if the time scale is very short.
The typical dynamical time scale of the collapsing core is $t_{\rm{dyn}} \sim 1/\sqrt{G \rho}$. 
As the result, the amplitude is roughly proportional to the core density $\rho_c$ multiplied by the radiative quadrupole moment: 
\begin{eqnarray}
h^{\rm TT} \propto \rho_c M^{E2}_{20}.
\label{propotion}
\end{eqnarray}
The leftsides of Figs.~\ref{fig:GWQh}-\ref{fig:GWQs} show the time dependence of the maximum density.
The maximum density of the model with the phase transition (for example the model of u-Mm)
 is always larger than that without the transition (for example the model of u-Bm). 
In the figures of $\rho_{max}$ and their corresponding GWA figures, there are high frequency spikes.
However, these are numerically ones, because 
quark matter areas ($\sim$ 10~km) are much smaller than overall calculational areas (about Fe-core size).
More fine tuned calculations with many meshes will delete such spikes.
The absolute values of $M^{E2}_{20}$ are almost the same around the peak,
 if the initial rotational strengths are the same (see the right panels of Figures~\ref{fig:GWQh}-\ref{fig:GWQs}). 
Since the transition causes the increase in the density being similar to the effect of soft EOS,
 we get larger amplitudes.
$|h^{\rm TT}|_{\textmd{max}}$ in u-Ms increases by 10\% compared to u-Bs as seen in Table \ref{tab:result}.
On the contrary, we find that the effect of the phase transition becomes small if the 
rotational strength becomes as large as $T/|W|_{\textmd{fin}}\simeq5\%$, 
which leads to the suppression of the contraction due to the strong centrifugal  forces (compare $\rho_{\textmd{max}}$ of u-Mm with that of U-Mm in Table \ref{tab:result}). 

For the Fourier transformations, the ranges of frequencies are distributed in 500-2000~Hz for all models. Clear peaks are not found
compared to the case in Ref.~\cite{lin}. 
While we adopt realistic presupernova models as initial models, they use polytropic models of equilibrium neutron stars. 
If QCD phase transition is considered, 
both the soft EOS model (FIG.~\ref{fig:GWs}) and slowly rotating model (FIG.~\ref{fig:GWm}) 
result in the shift to the high frequency sides in proportion to the increased density during the collapse (see $f_{1st.}$ of the models with/without the transition in Table \ref{tab:result}).

 To check the effects on GWA due to  the difference of presupernova models,
 we calculate the collapse of 40 $M_{\odot}$ models.
  The result is given in Table III (compare  model 40u-Mm with u-Mm, or 40d-Mm with d-Mm).
 We see the difference of the maximum density at the bounce, which is ascribed to
the size of the Fe-core. 
 The slight differences of the maximum amplitudes are found between the models with/without the phase transition. 
 The difference of the presupernova models dose not make a qualitative difference in both uniformly rotating
and differentially rotating models.
In the Fourier transformation, the overall frequencies for 40u-Mm shift to the higher region compared to the case
of 40u-Bm as seen in FIG.~\ref{fig:40u} except for around $f_{1st}$, where the first peak is not clear.

We change a value of coupling constant $\alpha_{s}$.
As the width of the density during the phase transition (coexistence area) becomes narrower for $\alpha_s \neq 0$,
the difference of the amplitude by the presence of the 
phase transition should  become small.
In Table~\ref{tab:result},
the value $|h^{TT}|_{max}$ of u-Mma is between u-Mm and u-Bm.
In the Fourier transformation, the change in $f_{1st}$ is small, but the frequencies shift overall between those of u-Mm and u-Bm
(see Table~\ref{tab:result} and FIG.~\ref{fig:GWua}).
In differentially rotating model, GWA and the frequencies of d-Mma lie between those of d-Mm and d-Bm, too.

The difference of critical density ($\rho_1$) for the gravitational wave is shown in Table~\ref{tab:result}.
The higher critical density~(u-Mm6) results in  the narrower width of the density region during the transition.
As a consequence, $|h^{TT}|_{max}$ and $f_{1st}$ of u-Mm6 are between the corresponding values of u-Mm and u-Bm.

 For the strong differential rotation, it is shown in 
Table~\ref{tab:result} and FIG.~\ref{fig:GWR} that the maximum amplitudes 
become smaller for the models 
with the phase transition 
(see $\Delta|h^{\textmd{TT}}|_{\textmd{max}}$ of the three models, 
from the bottom in table \ref{tab:result}). 
To explain such aspects, we refer to the early research~\cite{obergaulinger}.
Their results show that {\it very fast rotation with soft EOS} models tend to suppress the centrifugal force element 
of GWA. Here {\it very fast rotation} is not limited in differential rotation,
 but includes very fast uniform rotation whose $T/|W|$ is one digit larger than our uniformly rotating models.
Since our initial models are spherical models, we do not adopt such a large $T/|W|$ for consistency with our initial models.
Consequently, {\it the strong differentially rotating models with the transition} in our calculations are the same as {\it very fast rotation with soft EOS}, 
which lowers 1st peak of GWA for our differentially rotating models shown in Table~III. 
Corresponding to these, their frequencies with the transition shift to low
 (see their Fourier transformation, FIG.~\ref{fig:GWR} and $f_{1st.}$s of Table~\ref{tab:result}).
The above estimate (\ref{propotion}) is useful 
for the interpretation of the results. However it is found to be 
not applicable for strong differential rotation. This is because 
the differential rotation acts against a matter infall to the center by 
the phase transition due to the stronger centrifugal forces, but simultaneously 
leads to the stronger accretion to the center due to the smaller 
angular momentum in the rather outer regions. Due to this 
 very subtle competition, we can only know from the numerical results that the 
differential rotation makes the amplitudes lower up to $\sim $ 10 percents 
at the moment of the phase transition as far as the parameters in the present calculations.

\subsection{\label{sec:level2}Effects of the phase transition and magnetic fields}
\begin{figure}
\includegraphics[width=7.3cm]{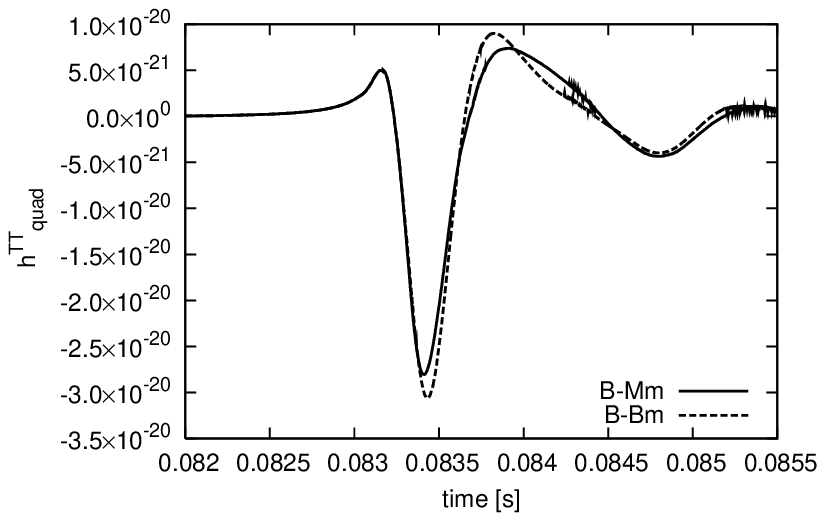}%
\\
\includegraphics[width=7.3cm]{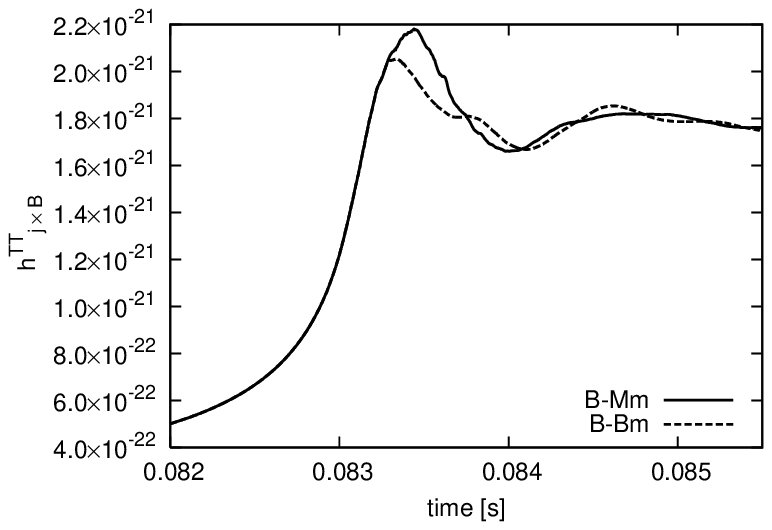}
\\
\includegraphics[width=7.3cm]{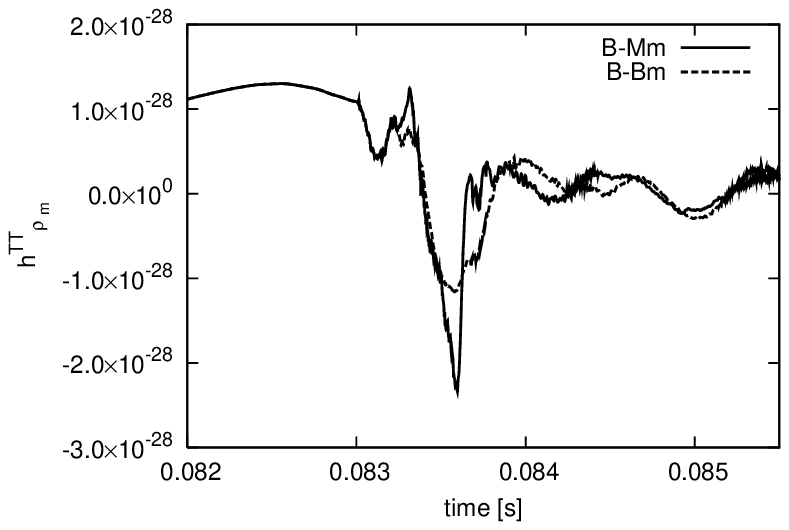}
\caption{\label{fig:AZZjB}   Waveforms of the strong magnetized models, B-Mm and B-Bm.
The top figure is one originated from the mass quadrupole moment, the next figure is from $\bm{j}\times \bm{B}$ part and the bottom is from the time
 derivatives of the magnetic energy density $\rho_m$.
Note the scale differences of the vertical axis.}
\end{figure}

We focus on the models with the strongest magnetic fields ("B$-$" models in Table \ref{tab:table01})
 to clarify the effects of the magnetic field on the gravitational wave. 
First, we compare each component $h^{\rm TT}_{\textmd{quad}}$, $h^{\rm TT}_{\bm{j} \times \bm{B}}$ and $h^{\rm TT}_{\rho_m}$
 of $h^{\rm TT}$~in Eq.(\ref{htt}).
Figure~\ref{fig:AZZjB} shows the waveforms originated from the mass quadrupole moment, $ \bm{j \times B}$ part,
 and the time derivative of the magnetic energy density $\rho_m$. 
The most definitive component to GWA is the one originated from the mass quadrupole moment $h^{\textmd{\rm TT}}_{\rm quad}$.

\begin{figure}
\includegraphics[width=7cm]{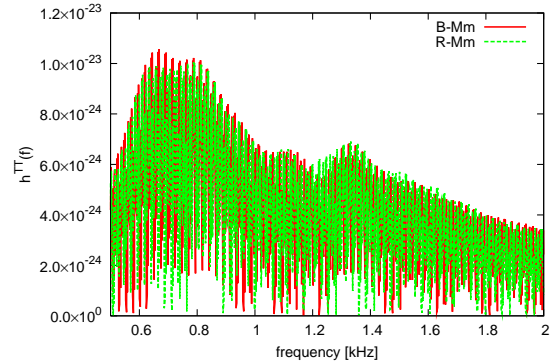}
\caption{\label{fig:Bfourier}  Fourier transformation of waveforms for the strongly magnetized model B-Mm and
unmagnetized model R-Mm, corresponding to GWAs of FIG.~\ref{fig:GWR} and FIG.~\ref{fig:AZZjB} .}
\end{figure}

To see the influence of the magnetic field on the gravitational wave frequencies,
we compare the zero magnetic field models (R-Bm, R-Mm) with strong magnetic field ones (B-Bm, B-Mm). 
It is understood from FIG~\ref{fig:GWR}-\ref{fig:Bfourier} that the influence of the magnetic field with the transition 
on the gravitational wave can be neglected.
The corresponding maximum GWAs have almost the same values as shown in Table~\ref{tab:result}. 

The components of GWA for B-Bm and B-Mm are summarized in Table \ref{tab:htt}. 
It should be noted that sign of each component at the bounce is different in the strong magnetic field models. 
As the result of cancellation between $h^{\textmd{\rm TT}}_{\textmd{quad}}$ and $h^{\textmd{\rm TT}}_{j \times B}$,
 the ratio of $|h^{\textmd{\rm TT}}_{j \times B}/h^{\textmd{\rm TT}}_{\textmd{quad}}|$ becomes 
 less than 8~\% regardless of the phase transition, which can be roughly 
estimated by the ratio of the magnetic to the matter energy density at bounce, 
namely,
\begin{equation}
\cfrac{B^2_c}{8 \pi } \cdot \rho_c^{-1} \sim O(1)~\% (\cfrac{B_c}{ 10^{17} \textmd{~G}})^2(\cfrac{\rho_c}{10^{14}\textmd{~g~cm}^{-3}})^{-1}.
\label{Rmag}
\end{equation}
Kotake et al. (2004) \cite{kotake} have found the same effect though they 
did not include the phase transition.
In fact, we can see from Table~\ref{tab:result} that GWAs of the model
B-Bm (B-Mm) is smaller compared to other models R-Bm (R-Mm) due to this 
 contribution from the magnetic fields.

At the moment of the phase transition, the magnetic fields in the central 
regions become larger due to the compression,  because the 
magnetic fields are frozen-in to the matter. As a result, the contribution 
to the amplitudes from the magnetic fields become larger for the 
model with the transition, but the change is found to be as small as
 1\% (compare $|h^{\textmd{\rm TT}}_{j \times B}/h^{\textmd{\rm TT}}_{\textmd{quad}}|$ in table \ref{tab:htt}.) 

\begin{table}
\caption{\label{tab:htt}Three components in Eq.~(\ref{htt}). Note that the listed values
correspond to the time
when the values of $|h^{\textmd{\rm TT}}_{j \times B}/h^{\textmd{\rm TT}}_{\textmd{quad}}|_{max}$ is maximum.
It means that the time is not same with the time of $|h^{\textmd{\rm TT}}|_{\textmd{max}}$ in table~\ref{tab:result}.
All amplitudes are given in units of $10^{-20}$.}
\begin{ruledtabular}
\begin{tabular}{lcccc}
Model & $h^{\textmd{\rm TT}}_{\textmd{quad}}$ & $h^{\textmd{\rm TT}}_{j \times B}$ &
 $h^{\textmd{\rm TT}}_{\rho_m}$  &  $|h^{\textmd{\rm TT}}_{j \times B}/h^{\textmd{\rm TT}}_{\textmd{quad}}|_{max}$(\%) \\
\hline
B-Bm & $-$3.06 & 1.99$\times 10^{-1}$ & $-$3.24 $\times 10^{-9}$ & 6.5 \\
B-Mm & $-$2.80 & 2.16$\times 10^{-1}$ & $-$3.61$\times 10^{-9}$ & 7.7 \\
\end{tabular}
\end{ruledtabular}
\end{table}
%

\section{Conclusion}
We have performed
two-dimensional axisymmetric, magnetohydrodynamical simulations for 
supernova cores accompanying the QCD phase transition. 
To elucidate the implications of a phase transition against a supernova,
 we investigated how the phase transition affects the gravitational waveforms 
near the epoch of core-bounce. As for the initial models, we changed parametrically
 the strength of the rotation and the magnetic fields. As for 
the microphysics, we adopted a phenomenological equation of state above 
the nuclear matter density, including two parameters to change the 
hardness of the matter before the transition. 
To treat the QCD phase transition, we employed a MIT bag model.
Based on these computations, 
we showed that the phase transition can make the maximum amplitudes
 larger up to $\sim$ 10 percents than the ones without the phase transition.
On the other hand, we found that the phase transition makes the maximum 
amplitudes smaller up to $\sim$ 10 percents, when the iron core rotates strongly differentially. It was confirmed that the strong magnetic fields themselves  
decrease the maximum amplitudes by less than 8\% regardless of the transition.
Even the extremely strong magnetic fields $\sim 
10^{17}$ G in the protoneutron star do not affect the above features.

Finally, we give some discussions and speculations based on the obtained 
results. We assume rather low density for the onset of 
the phase transition in comparison with a recently predicted EOS by the  
 lattice QCD computations \cite{ivanov}. 
Thus the results here could be some extreme cases 
predicted by the phenomenological MIT bag model, and the change in the 
amplitudes due to the transition could be interpreted as an upper bound.

Although the other initial presupernova models may change our analysis 
to some extent, the qualitative results obtained here will not 
be different so much. 
This is because the structure of the iron core is similar, while 
 the initial mass of the helium core increases with the progenitor 
masses \cite{hashimoto}. 
For the presupernova model of $40 M_\odot$ 
that has the iron core of 1.9$M_\odot$,  we have found that GWA at the bounce
is only about 3 percents larger than that of 13$M_{\odot}$.
Since we concentrate on the forms of the gravitational wave just after the bounce, the difference in the helium core mass should not be important at all.

 If we could calculate the hydrodynamical evolution long after the core 
bounce especially in some failed core-collapse supernova models, 
we would observe the phase-transition of the protoneutron stars.
We also think interesting to investigate the possible effects of 
color-superconductivity recently proposed \cite{keranen}. 
As a next step, we are now going to investigate how the gravitational 
waves will be originated from such events in the context of 
magnetorotational core-collapse. 

\begin{acknowledgments}
N.Y. is grateful to H. Ono \& H. Suzuki for their helpful instruction about EOS,
 and N. Nishimura for the discussion of the magneto-hydrodynamical simulations.
This work was supported in part by the Japan Society for
Promotion of Science(JSPS) Research Fellowships (N.Y.), 
Grants-in-Aid for the Scientific Research from the Ministry of
Education, Science and Culture of Japan (No.S14102004, No.14079202, 
No.17540267), and Grant-in-Aid for the 21st century COE program
``Holistic Research and Education Center for Physics of
Self-organizing Systems''
\end{acknowledgments}

\newpage 

\end{document}